\documentclass{raa}
\usepackage{graphicx,times}
\usepackage{natbib}
\usepackage{amssymb,amsmath}
\bibpunct{(}{)}{;}{a}{}{,}
\usepackage[a4paper=true,dvipdfm=true,pagebackref=true]{hyperref}
\hypersetup{pdftitle = The title of my PDF, pdfauthor = My name, pdfsubject= The subject, pdfkeywords = keyword1 keyword2 keyword3}
\hypersetup{colorlinks = true, linkcolor = green, anchorcolor = red, citecolor = blue, filecolor = red, pagecolor = red, urlcolor = red}
\makeatletter
\newcommand*{\rom}[1]{\expandafter\@slowromancap\romannumeral #1@}
\makeatother

\def\arcsec{\hbox{$^{\prime\prime}$}}

\def\sun{\hbox{$\odot$}}                    
\def\degr{\hbox{$^\circ$}}

\begin{document}%

\title{Physical and Geometrical Parameters of VCBS~\rom{13}: HIP\,105947}

\volnopage{ {\bf 2018} Vol.\ {\bf 18} No. {\bf 3}, 000--000}
   \setcounter{page}{1}

   \author{ Suhail G. Masda\inst{1,2}, Mashhoor A. Al-Wardat\inst{3} and J. M. Pathan \inst{4}}

\institute{  Physics Department, Dr. Babasaheb Ambedkar Marathwada University, Aurangabad-431001, Maharashtra, India; \textit{suhail.masda@gmail.com} \\
		 \and Physics Department, Hadhramout University, PO Box:50511, Mukalla, Yemen \\
	     \and Dep. of Physics and Institute of Astronomy and Space Sciences, Al al-Bayt University, PO Box: 130040, Mafraq, 25113 Jordan\\
	     \and Physics Department, Maulana Azad College, Aurangabad-431001, Maharashtra, India\\
\vs \no
   {\small Received 2017 Sep 24; Accepted 2018 Feb 8}
}

\abstract{
The perfect physical and geometrical parameters of the main-sequence close visual binary system (CVBS); HIP\,105947, are explicitly presented. These parameters have been constructed conclusively  using Al-Wardat's complex method for analysing CVBSs, which is a method for constructing a synthetic  spectral energy distribution (SED) for the entire binary system using individual SED for each component star. Which in its turn built using Kurucz ({\fontfamily{cmtt}\selectfont ATLAS9}) line-blanketed plane-parallel models.
At the same time, the orbital parameters for the system are calculated  using Tokovinin's dynamical method for constructing the best orbits of interferometric binary system. Henceforth, the mass-sum of the components, the $\triangle\theta $ and $\triangle\rho $ residuals for the system are introduced. The combination of Al-Wardat's and Tokovinin's methods yields the estimation of the physical and geometrical parameters to their best values. The positions of the components of the system  on the evolutionary tracks and isochrones are plotted and their formation and evolution are discussed.
}

\authorrunning{Masda et al. }            
   \titlerunning{Parameters of CVBS ~\rom{13}: HIP\,105947}  
\maketitle%

\section{INTRODUCTION }
 Studies of duplicity and multiplicity of stars in the Galaxy showed that their ratio  reaches  50 per cent among  nearby solar-type main-sequence stars  ~\cite{1991A&A...248..485D}, and 42 per cent  among nearby M stars ~\cite{1992ApJ...396..178F}. Most of these stars are either visually close enough to each other or  far away that they appear as single stars, i.e. the separation angle between the components  is small, and henceforth will be called close visual binary stars (CVBSs).

 One of the \textit{Hipparcos} mission goals was to reveal the duplicity of these stars and it reported thousands of resolved stars and their relative position measurements. At the same time and later on, several astronomical groups used special techniques like speckle-interferometry and  adaptive optics with ground-based telescopes to observe these stars and to measure their relative positions and magnitude differences.

 The study and analysis of stellar binary systems is an extremely sensational and productive topic, especially for the estimation of stellar masses and distances which is usually
 derived in a considerable way from accurate analysis of  synthetic spectra, keplerian motions, radial velocity curves and theoretical evolutionary tracks ~\citep{2003Ap&SS.283..309D,2010MNRAS.401..695M}.

 Some of the physical and geometrical parameters can be estimated directly in the cases of eclipsing and spectroscopic binaries, while there is no direct way to estimate the parameters of CVBSs ~\citep{2017AstBu..72...24A}. So, Al-Wardat's complex method for analysing CVBSs ~\citep{2016RAA....16..166A} presents an indirect method which combines magnitude difference measurements from speckle interferometry with the system's entire spectral energy distributions from  spectrophotometry along with the atmospheric modelling to estimate the physical and geometrical parameters, and it was combined with Tokovinin's method for dynamical analysis ~\citep{1992ASPC...32..573T}  to get the complete set of these parameters for the individual components of a binary system.

 The method has been applied with substantial success to several  solar-type stars like ADS\,11061, COU\,1289, COU\,1291, HIP\,11352, HIP\,11253, HIP\,70973, HIP\,72479,  Gliese\,150.2, Gliese\,762.1, COU\,1511, and FIN\,350 ~\citep{2002BSAO...53...51A, 2007AN....328...63A, 2009AN....330..385A, 2009AstBu..64..365A, 2012PASA...29..523A, 2014AstBu..69..198A,2016RAA....16..166A,2016RAA....16..112M,2017AstBu..72...24A} (henceforth papers I, II, III, IV, V, VIII, X, XI and  XII, respectively)  and to a most rare instances like  sub-giant binary systems HD\,25811,  HD\,375 and HD\,6009 ~\cite{2014PASA...31....5A,2014AstBu..69...58A, 2014AstBu..69..454A} (henceforth paper VI, VII and IX in this series respectively).

 This paper, the 13th of a series, presents the complete collection of the physical and geometrical parameters of the solar-type close visual binary system; HIP\,105947 (HD\,204236).

 As a matter of fact, HIP\,105947 is known as solar neighborhood (CVBS) located at a kinematic distance of $ 57.21\pm0.003$\,pc ($17.48\pm1.02$\,mas)~\cite{2007A&A...474..653V} and its measured separation is less than or equal to $0\arcsec.2$. The system was detected through the \textit{Hipparcos} mission and  the first observation was given for the 1991.25 epoch of the Hipparcos catalog~\cite{1997yCat.1239....0E}.

 The first orbit of HIP\,105947  was calculated  by ~\citep{2006AA...448..703B} depending on  \cite{1977ApJ...214L.133M} method, their estimated mass-sum was $2.68\pm0.58\rm\, M_\odot$  using the old \textit{Hipparcos} parallax of $17.11\pm1.13$\,mas~\citep{1997yCat.1239....0E}. After that it was modified by ~\citep{2010AJ....140..735M} depending on the methods described by \citep{2001AJ....122.3480H}, their estimated mass-sum was $2.30\pm1.14\rm\,M_\odot$ using the revised \textit{Hipparcos} parallax of  $17.48\pm1.02$\,mas~\citep{2007A&A...474..653V}.
 The last observational relative position measurement used by ~\cite{2010AJ....140..735M}  was at epoch 2009.6707. So, five new observational points from epoch 2010.4818 to epoch 2015.4974 are included in our modified orbit (Table~\ref{tab}).
 \begin{table}[h]
 	\centering
 	\caption{New data of Interferometric Measurements for the HIP\,105947 system.} \label{tab}
 	\begin{tabular}{ccccccc}\hline\hline
 		& Data  & $\theta$ & $\rho$ & ($\lambda/\Delta\lambda$)& Tel.$^*$& References
 		\\
 		& Epoch  & (\degr)& ($''$) & (nm) & m & \\
 		\hline
 		HIP\,105947  & 2010.4818 & 165.5 & 0.1788 & 562/40 & 3.5 & \citep{2011AJ....141...45H}\\
 		& 2013.7363 & 225.8 & 0.0850 & 534 /22& 4.1 &\citep{2014AJ....147..123T}\\
 		& 2014.7629  & 266.6 & 0.0726 & 778/132 & 4.2 &\citep{2015AJ....150...50T}\\
 		& 2015.4974  & 304.4 & 0.0908 & 543/22 & 4.1 & \citep{2016AJ....151..153T}\\
 		& 2015.4974  & 304.8 & 0.0952 & 788/132 & 4.1 &\citep{2016AJ....151..153T}\\
 		\hline\hline
 	\end{tabular}
 	\\
 	$^*$ Telescope aperture in meters where the
 	observations were obtained.
 \end{table}
\section{Observational data}
The entire observational spectral energy distribution (SED) for the system; HIP\,105947 was taken from~\cite{2003BSAO...55...18A} (Figure~\ref{aa00}), and it was used as reference for the comparison with entire synthetic SED. The observational SED (Figure~\ref{aa00}) was  obtained  using  a low-resolution grating ($325/4^{\circ}$ grooves/mm, $5.97$ {\AA}/px reciprocal dispersion) within the UAGS spectrograph at the 1\,m (Zeiss-1000) telescope of SAO-Russia and it covers the approximate wavelength range $\lambda$ 3700 to~ 8000\,\AA.

Table~\ref{tab1} contains the fundamental data for  HIP\,105947. These data were taken  from SIMBAD database, NASA/IPAC, Hipparcos, and Tycho Catalogues ~\cite{1997yCat.1239....0E} and Str\"{o}mgren and Table~\ref{tab22} shows the magnitudes difference $\rm \Delta m$ between the components of the
system along with filters used in the observation expressed in nm and reference for each value from Fourth Catalog of Interferometric Measurements of Binary Stars (INT4)~\footnote{http://www.usno.navy.mil/USNO/astrometry/optical-IR-prod/wds/int4}.
\begin{table}[h]
	\centering
	\caption{The real observations data from SIMBAD database and real magnitudes and colour indices from Hipparcos, and Tycho Catalogues and Str\"{o}mgren for the HIP\,105947 system.} \label{tab1}
	\begin{tabular}{ccc}\hline\hline
		Property&  HIP\,105947 & Source of data 		\\
		& HD\,204236 &  \\
		\hline
		$\alpha_{2000}$ $^1$&   $21^{\rm h} 27^{\rm m} 23^{\rm s}255$& SIMBAD$^2$\\
		$\delta_{2000}$ $^3$ 	&$-07\degr00' 56.''95$ & -\\
		SAO  & 145409 &  -\\
		Sp. Typ.  &  F8 &-
		\\
		E(B-V)      & $0.10\pm0.003$ & NASA/IPAC$^{4}$
		\\		
		$A_V$       &  $0.31\pm 0.003$ & -
		\\
		$V_J(Hip)$  & $7.53$ &  \cite{1997yCat.1239....0E}
		\\
		$B_J$       &  $8.10\pm0.02$  & $5$
		\\
		$(V-I)_J$&   $0.67\pm0.01$ & \cite{1997yCat.1239....0E}
		\\				
		$(B-V)_J$&   $0.60\pm0.02$ & -
		\\
		$B_T$  &  $8.24\pm0.02$ & 5
		\\			
		$V_T$ &   $7.59\pm0.01$ & 5
		\\
		$(u-v)_S$& $0.95\pm0.02$ &  6\\
		$(v-b)_S$& $0.56\pm0.01$ &  6\\
		$(b-y)_S$& $0.37\pm0.00$ &  6
		\\
		$\pi_{Hip}$ (mas)      & $17.11\pm1.13$ & \cite{1997yCat.1239....0E}
		\\
		$\pi_{Hip}$ (mas)    &  $17.48\pm1.02$ & \cite{van2007validation}
		\\
		\hline\hline
	\end{tabular}
	\\
	Notes.\\
	$^1$ Right Ascention,  	$^2$ http://simbad.u-strasbg.fr/simbad/sim-fid.\\
	$^3$ Declination,
	$^{4}$~\citep{1998ApJ...500..525S},
	$^5$~\citep{2000A&A...355L..27H},
	$^6$~\citep{1998A&AS..129..431H},
\end{table}
\begin{table}[h]
	\begin{center}
		\caption{Magnitude difference between the components of the
			HIP\,105947 system and available errors , along with filters used to obtain the observations. }
		\label{tab22}
		\begin{tabular}{cccc}
			\noalign{\smallskip}
			\hline\hline
			$\triangle\rm m $& {$\sigma_{\Delta\rm m}$}& Filter ($\lambda/\Delta\lambda$)&  Ref.$^{**}$  \\
			(mag) & & ($~\rm nm$)&    \\
			\hline
			$1.17$ &   0.60  & $V_{Hip}:550/40 $&  1   \\				
			$1.50$ &  0.04   & $545/30 $& 2     \\		
			$1.40$ &   0.02  &$545/30 $& 3      \\
			$1.29$ &  0.05  & $850/75$ & 3    \\		
			$1.31$ &  0.06  &$600/30 $ &  3   \\
			$1.47$ &  0.03  &$545/30 $& 4   \\			
			$1.56$ & *     & $550/40 $  &  5    \\
			$1.56$ &  *    & $550/40 $  & 6    \\
			$2.00$ &  *   & $551/22 $  &  7       \\
			$1.40$ &   *  & $657/05 $ &  7   \\	
			$1.50$ &  *   & $551/22 $  &  7         \\
			$1.50$ &   *   & $788/132 $  & 8         \\
			$1.37$ &   *   & $692/40 $  &  9       \\
			$1.44$ &   *  & $562/40 $  &  9       \\
			$1.90$ &   *  & $534/22 $  &  10        \\
			$1.10$ &   *  & $788/132 $  &  11       \\
			$1.30$ &   *  & $543/22 $  &  12        \\
			$1.50$ &   *  & $788/132 $  &  12       \\
			\hline\hline
		\end{tabular}
		\\
		$^*$ The errors are not given in Fourth Catalog of Interferometric Measurements of Binary Stars (INT4).\\		
		$^{**}$	$^1${\cite{1997yCat.1239....0E}},
		$^2${\cite{2005A&A...431..587P}}
		$^3${\citep{2006AA...448..703B}},
		$^4${\citep{2007AstBu..62..339B}}, 		
		$^5${\citep{2008AJ....136..312H}}, 		
		$^6${\citep{2010AJ....139..205H}}, 		
		$^7${\citep{2010AJ....139..743T}}, 		
		$^8${\citep{2010PASP..122.1483T}}, 		
		$^9${\citep{2011AJ....141...45H}},
		$^{10}${\citep{2014AJ....147..123T}},
		$^{11}${\citep{2015AJ....150...50T}},
		$^{12}${\citep{2016AJ....151..153T}}.
	\end{center}
\end{table}

\section{Method and Analysis}

\subsection{HIP\,105947}\label{22}
Entire visual magnitude of the binary ($\rm m_v$ $(V_J)$) and magnitude difference between its components ($\triangle\rm m$)  are the start keys of applying Al-Wardat's complex method for analysing CVBSs. Together, they led to the determination of  the apparent magnitudes of the individual components of the system  using the following simple relationships:
\begin{eqnarray}\centering
\label{eq111}
\rm m_v^A=\rm m_v+2.5\log(1+10^{-0.4\triangle\rm m}),
\end{eqnarray}
\begin{eqnarray}
\centering
\label{eq222}
\rm m_v^B=\rm m_v^A+{\triangle\rm m},
\end{eqnarray}

\noindent and their errors  follow using the following equations:
\begin{eqnarray}
\label{eq333}
\sigma^2_{\rm m^A_{v}} = {\sigma_{\rm m_{v}}^2+(\frac{1}{1+10^{+0.4\triangle\rm m}})^2\sigma_{\triangle\rm m}^2},
\end{eqnarray}
\begin{eqnarray}
\label{eq3311}
\sigma^2_{\rm m^B_{v}} = {\sigma_{\rm m^A_{v}}^2+\sigma_{\triangle\rm m}^2}
\end{eqnarray}

For the system HIP\,105947, the mean speckle-interferometric magnitude difference was taken as $1^{\rm m}.53\pm0.07$, which is the average of all $\triangle\rm m$ measurements under different speckle filters for V-band  filters (534 -562) nm with different band widths  (see Table~\ref{tab22}), as the closest filters to $V_J$.

This value was implemented concurrently with the entire visual magnitude $\rm m_v=7^{\rm m}.53$ of the system (Table~\ref{tab1}) in Equations~\ref{eq111}~\&~\ref{eq222}, and resulted  in calculating the individual apparent magnitudes of the  components of the system as $\rm m_v^A=7^{\rm m}.77\pm0.01 $ and $\rm m_v^B=9^{\rm m}.30\pm0.07$ for the primary and secondary components of the system, respectively.

Using  the updated \textit{Hipparcos} parallax of $\rm d=57.21\pm0.003$\,pc
\cite{2007A&A...474..653V}, with the following relation:
\begin{eqnarray}
\label{eq3}
\rm M_V-m_v=5-5\log(d)-A_V,
\end{eqnarray}
the absolute magnitudes along with their errors  follow as: $\rm M_V^A=3^{\rm m}.98\pm0.13$ and $\rm M_V^B=5^{\rm m}.51\pm0.14$. Note that the interstellar extinction can be neglected ($A_V\approx0$) because the system is a nearby one.

The errors of the absolute magnitudes were calculated using the following relation:
\begin{eqnarray}
\label{eq341}
\sigma^2_{\rm M^{*}_{V}} ={\sigma_{\rm m^{*}_{v}}^2+(\frac{ \log\rm e}{0.2\pi_{Hip}})^2\sigma_{\pi_{Hip}}^2} ~;~~~~~ \textit{*}=A,B.
\end{eqnarray}
\noindent where  $ \sigma_{\rm m^*_{v}} $ are the errors of the apparent magnitudes of the A and B components  in Equations~\ref{eq333}~\&~\ref{eq3311}.

Hence, depending on the estimated preliminary absolute magnitudes,  the  preliminary values of the effective temperature and gravity acceleration for each component were taken from the Tables of \cite{1992adps.book.....L} and \cite{2005oasp.book.....G} as follows: $T_{\rm eff}=6200\,\rm K$, log g = 4.31  for the primary component and  $T_{\rm eff}=5490\,\rm K$, log g = 4.47  for the secondary component.

Note that for solar-type stars, these parameters are related to other parameters according to the following equations:
\begin{eqnarray}
\label{eq81}
\log\frac{R}{R_\odot}= \frac{M_{bol}^\odot-M_{bol}}{5}-2\log\frac{T_{\rm eff}}{T_\odot},\\
\label{eq82}
\log g = \log\frac{M}{M_\odot}- 2\log\frac{R}{R_\odot} + \log g_\odot.
\end{eqnarray}	
\noindent where $T_\odot=5777\,\rm K$, $M_{bol}^\odot=4^{\rm m}.75$ and log $ g_\odot=4.44$. $M_{bol}=\rm M_{V}+BC$; $\rm BC$ indicates the bolometric correction.

Now, we have the preliminary input parameters to build model atmospheres for each individual component of the system, and to do so we use  {\fontfamily{cmtt}\selectfont ATLAS9} line-blanketed model atmospheres of ~\cite{1994KurCD..19.....K}.

In order to build a synthetic SED of a star from its model atmospheres, we need information about its distance d and  radius R. The first parameter d was taken from the updated \textit{Hipparcos} parallax  ($\rm d=57.21\pm0.003$\,pc)	as a postulate value with its error, while the second parameter is subject to change within the values of the observational data like entire SED, magnitudes and colour indices.	

The method make use of the entire observational SED as a reference guide for the best-fitted entire synthetic SED,  in an iterated way.
The entire synthetic SED of the binary system, which is connected to the energy flux of the individual components A and B  located at a distance d\,(pc) from the Earth, is calculated using the following equation:
\begin{eqnarray}
\label{eq66}
F_\lambda \cdot d^2 = H_\lambda ^A \cdot R_{A} ^2 + H_\lambda ^B
\cdot R_{B} ^2,
\end{eqnarray}
\noindent This equation can be written as
\begin{eqnarray}
\label{eq77}
F_\lambda  = (\frac{R_{A}}{d})^2(H_\lambda ^A + H_\lambda ^B (\frac{R_{B}}{R_{A}})^2) ,
\end{eqnarray}

\noindent
where $F_\lambda$ is the flux for the entire synthetic  SED of the entire binary system at the Earth, $H_\lambda ^A $ and  $H_\lambda ^B$ are the fluxes of the primary and secondary  star, in units of ergs cm$^{-2}$\rm s$^{-1}$ \AA$^{-1}$, while $ R_{A}$ and $ R_{B}$ are the radii of the primary and secondary components of the binary system in solar units.

Many attempts were carried out tell the best-fit between the synthetic and observational SEDs was achieved. Hence, the physical and geometrical  parameters of the best-fitted synthetic SED represent adequately enough the parameters of the system's components within the error limits of  the observational SED.

It is very important to mention here that  the values of the radii are totally reliant on \textit{Hipparcos} parallax measurements precision (see Equation ~\ref{eq77}), which in some cases is distorted by the orbital motion of the components of such systems as pointed out by \cite{1998AstL...24..673S}.

\begin{figure}[h]
	\centering
	\includegraphics[angle=0,width=12cm]{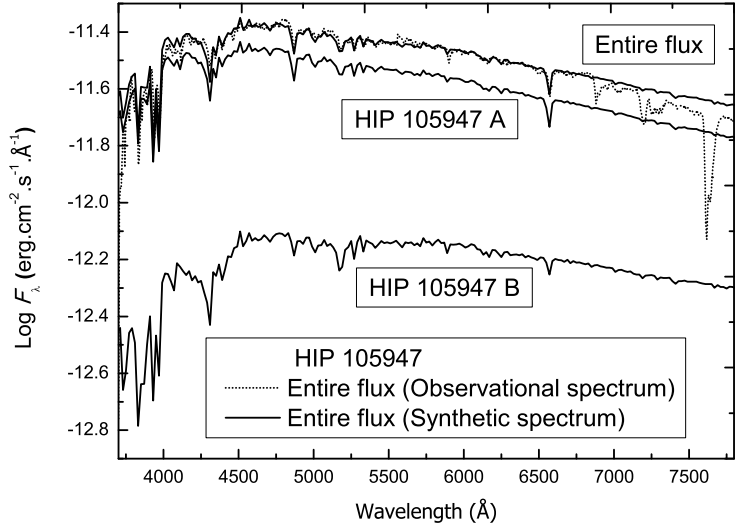}
	\caption{The entire synthetic SED of the system HIP\,105947 (built using  parameters given in Table ~\ref{tablef1} like $T_{\rm eff}$, $\log \rm g$, R and Parallax) against the observational one, the Figure also shows the synthetic SED for each component.} \label{aa00}
\end{figure}

The parameters which led to the best-fit were as follows (Figure~\ref{aa00}):
$$T_{\rm eff}^{A}=6230\pm80\,\rm K, T_{\rm eff}^{B} =5470\pm80\,\rm K,$$
$$ \log \rm g_A=4.30\pm0.07, \log\rm g_B=4.45\pm0.08$$
$$R_{A}=1.443\pm0.06\, R_\odot, R_{B}=0.989\pm0.05\,\rm R_\odot$$

The errors of the radii were estimated from the method and double-checked using the following equation:

\begin{eqnarray}
\label{eq3412}
\sigma_{R} \approx\pm R\sqrt{(\frac{\sigma_{M_{\rm bol}}}{5 \log e})^2+4(\frac{\sigma_{T_{\rm eff}}}{ T_{\rm eff} })^2}
\end{eqnarray}

The individual  stellar luminosities and bolometric magnitudes of the system follow as:
$L_A=2.816\pm0.30\,\rm L_\odot$, $L_B=0.786\pm0.10\,\rm L_\odot$,
$M^{A}_{bol}=3^{\rm m}.63\pm0.08$ and $M^{B}_{bol}=5^{\rm m}.01\pm0.09$, respectively.

Figures~\ref{evolall}\&\ref{evolall12} show  the positions of the components on the Hertzsprung-Russell (H-R) evolutionary tracks and isochrones of the binary system. Depending on that, the masses of the components can be estimated as:
$M^A=1.21\pm0.21\,\rm M_\odot$, $M^B=0.89\pm0.15\,\rm M_\odot$ with spectral types F8 for the primary and G7.5 for the secondary and  the age of the system as  $\rm t_T\approx 2.75\pm0.06$\,Gyr.

\section{Synthetic photometry}
Notwithstanding the significance and easiness  of the direct comparison between synthetic and observed spectra in determining the best-fit,  there is  another powerful and quantitative technique which examines this fitting, it is the synthetic photometry. It helps us to calculate the apparent magnitudes and colour indices of the entire and individual SEDs  in any photometrical system, and enables us to compare the synthetic photometry with the observational one.

The apparent magnitude for a specific photometric filter can  be done using the following equation ~\citep{2007ASPC..364..227M,2012PASA...29..523A}:

\begin{equation}\label{55}
m_p[F_{\lambda,s}(\lambda)] = -2.5 \log \frac{\int P_{p}(\lambda)F_{\lambda,s}(\lambda)\lambda{\rm d}\lambda}{\int P_{p}(\lambda)F_{\lambda,r}(\lambda)\lambda{\rm d}\lambda}+ {\rm ZP}_p\,
\end{equation}
\noindent	where $m_p$ is the synthetic magnitude of the passband $p$, $P_p(\lambda)$ is the dimensionless sensitivity function of the passband $p$, $F_{\lambda,s}(\lambda)$ is the synthetic SED of the object and $F_{\lambda,r}(\lambda)$ is the SED of the reference star (Vega).  Zero points (ZP$_p$) from~\cite{2007ASPC..364..227M} are adopted.

The  calculated magnitudes and colour  indices of the individual components and entire synthetic SEDs for the system, in different photometrical systems (Johnson: $U$, $B$, $ V$, $R$, $U-B$, $B-V$, $V-R$; Str\"{o}mgren: $u$, $v$, $b$,
$y$, $u-v$, $v-b$, $b-y$ and Tycho: $B_{T}$, $ V_{T}$, $B_{T}-V_{T}$), are listed  in Table~\ref{tab41}.

Table ~\ref{synth3} shows a good agreement between synthetic and observational magnitudes and colour indices, which reflects well on the best-fit between the synthetic and observational SED and hence the reliability of Al-Wardat's complex method in estimating the physical and geometrical parameters of CVBS.
\begin{table}[h]
	\small
	\begin{center}
		\caption{ Magnitudes and colour indices  of the entire synthetic spectrum and individual components of  HIP\,105947.}
		\label{tab41}
		\begin{tabular}{ccccc}
			\noalign{\smallskip}
			\hline\hline
			\noalign{\smallskip}
			Sys. & Filter & Entire Synth. & HIP\,105947 & HIP\,105947 \\
			&     & $\sigma=\pm0.03$&   A    &     B      \\
			\hline
			\noalign{\smallskip}
			Joh-          & $U$ & 8.24 & 8.39 & 10.47 \\
			Cou.          & $B$ & 8.13   &  8.32 &  10.10  \\
			& $V$ & 7.53 &  7.77 &  9.30 \\
			& $R$ & 7.20  &  7.46 & 8.88  \\
			&$U-B$& 0.11  & 0.07 & 0.37 \\
			&$B-V$& 0.60  &  0.55 &  0.79 \\
			&$V-R$& 0.33  &  0.30 & 0.42 \\
			\hline
			\noalign{\smallskip}
			Str\"{o}m.    & $u$ & 9.40 & 9.56 &  11.61  \\
			& $v$ & 8.45 & 8.63  & 10.52  \\
			& $b$ & 7.87 & 8.08 &  9.73 \\
			&  $y$& 7.50 & 7.74 & 9.27  \\
			&$u-v$& 0.95 & 0.93 & 1.09 \\
			&$v-b$& 0.59 & 0.55 & 0.79 \\
			&$b-y$& 0.37 & 0.34& 0.47 \\
			\hline
			\noalign{\smallskip}
			Tycho       &$B_T$  & 8.27   & 8.45 & 10.31   \\
			&$V_T$  & 7.60   & 7.83& 9.39  \\
			&$B_T-V_T$& 0.67 & 0.62 & 0.92\\
			\hline\hline
			\noalign{\smallskip}
		\end{tabular}
	\end{center}
\end{table}

\section{Orbits and masses}
\subsection{Orbits}

The orbital solution also plays a vital role in calculating  stellar masses. This solution involves the orbital period P (in years); the eccentricity e; the semi-major axis a (in arcsec); the inclination $ \rm i$ (in degree); the argument of periastron $ \rm \omega$ (in degree); the position angle of nodes $\rm \Omega$ (in degree); and the time of primary minimum $\rm T_0$ (in years).

In order to derive the orbital parameters for the system, we followed Tokovinin's  dynamical method \cite{1992ASPC...32..573T}. The method  performs a least-squares adjustment to all available  relative position measurements and radial velocity (once available), with weights inversely proportional to the square of their standard errors.

The initial determinations of the orbital parameters were carried out based on four parameters (P, T, e and a). These parameters are intrinsic
keys in applying the least-squares method and then to get the rest of the orbital parameters.

\begin{table}
	\begin{center}
		\centering
		\caption{ Relative position measurements, residuals $\triangle\theta $ and $\triangle\rho $ (Our work) and rms for HIP\,105947, which are used to build the orbit of the system. $\theta$ and $\rho$ are taken from the Fourth Catalog of Interferometric Measurements of Binary Stars. }
		\small
		\label{po1}
		\centering
		\begin{tabular}{lrcccr}
			\hline\hline
			Epoch	& $\theta$ & $\rho$& Ref. & $\triangle\theta $  & $\triangle\rho $   \\
			
			&(\degr)    &  ($''$)  &     &  (\degr) & ($''$)
			\\
			\hline
			\centering
			1991.25 & $171.64$  & $0.1266$&  1  & -11.1 & -0.008  \\
			1997.7201  & 177.6 $^*$ & $0.151$ & 2 & - & -	  \\
			1998.7796  & $219.2$$^{**}$ & $0.108$ & 3 & 5.7 & -0.001 \\
			1999.7468 & $242.2$$^{**}$ &  $0.110$ & 3 & 0.3  & -0.002 \\
			2000.8643 & $84.6$&  $0.127$  & 4  & 0.1 & -0.000 \\
			
			2001.7523 & $98.5$  &  $0.144$ & 4 & 0.3 & -0.000   \\
			2001.7523 & $98.5$  &$0.143$  & 4 & 0.3 & -0.001   \\
			
			2002.7986 & $112.2$  &  $0.170$ &  5& 1.6&   0.005 \\
			2002.7987 & $110.9$  & $0.168$&  5 & 0.3 & 0.003  \\
			2002.799 & $110.0$  & $0.163$  & 2& -0.6 & -0.002  \\
			2002.799 & $110.7$  & $0.163$  & 2& 0.1 &  -0.002  \\
			
			2004.815 & $127.5$  & $0.195$  &  2 & -0.2 & -0.004  \\
			2004.8152 & $ 127.5$  & $0.195$ & 6  & -0.3 &  -0.004   \\
			
			2006.4436  & $138.2$  & $0.211$ &  5 & -0.2 & -0.003			\\
			2006.5254 & $139.0$  & $0.226$ & 7 & 0.1 &  0.011   \\
			2006.6870 & $140.7$ &  $ 0.215$ &  5 & 0.8 & -0.000  \\

			2008.4613 & $151.5$ &   $0.214 $ & 8 & 0.7 & 0.003 \\
			2008.5430 & $151.5$ &   $0.2102$ & 9 & 0.2& -0.001  \\
			2008.7669 &$152.6$ &  $0.2084$  & 9 &-0.1 &  -0.000  \\
			2008.7669 &$ 152.5$  &$ 0.2073$ & 9 & -0.2&  -0.001 \\
			
			2009.6706 & $ 159.0$ &  $ 0.1951$  & 10 & 0.0 & -0.001\\
			2009.6707 & $159.0$ &   $0.195  $ & 11 & 0.0 & -0.001  \\
			
			2010.4818 $^{***}$ & $165.5$ &   $ 0.1788$ & 12 & 0.1 &  -0.001 \\
			2013.7363 $^{***}$ &$225.8 $ &  $0.0850$ & 13 & -2.5 & 0.003   \\
			
			2014.7629 $^{***}$&$266.6$  &$0.0726$  & 14 & -10.1  & -0.006 \\
			2015.4974 $^{***}$&$304.4$  &$0.0908$  & 15 & -0.4  & -0.004 \\
			2015.4974 $^{***}$&$304.8$  &$0.0952$  & 15 & 0.0  & 0.001 \\
			\hline
			rms &  &    &   & 0.60  & 0.0015  \\
			\hline\hline
		\end{tabular}
		\smallskip\\
		Notes:\\
		$^*$ It was neglected in our orbits~\citep{1999AJ....117.1890M},
		$^{**}$ These points were modified by $180\degr$ to achieve consistency with nearby points and $^{***}$ New observations in our orbit.\\
		$^1$    {\cite{1997yCat.1239....0E}},
		$^2$	{\citep{2006AA...448..703B}},
		$^3$	{\citep{2002A&A...385...87B}},
		$^4$	{\citep{2002AstL...28..773B}},
		$^5$	{\citep{2013AstBu..68...53B}},
		$^6$	{\citep{2007AstBu..62..339B}},
		$^7$	{\citep{2008AJ....136..312H}},
		$^8$	{\citep{2010AJ....139..205H}},
		$^9$	{\citep{2010AJ....139..743T}},
		$^{10}$	{\citep{2010PASP..122.1483T}},
		$^{11}$	{\citep{2010AJ....140..735M}},
		$^{12}$	{\citep{2011AJ....141...45H}},
		$^{13}$	{\citep{2014AJ....147..123T}},
		$^{14}$	{\citep{2015AJ....150...50T}}.
	\end{center}
\end{table}

\begin{figure}[ht]
	\centering
	\includegraphics[angle=0,width=12cm]{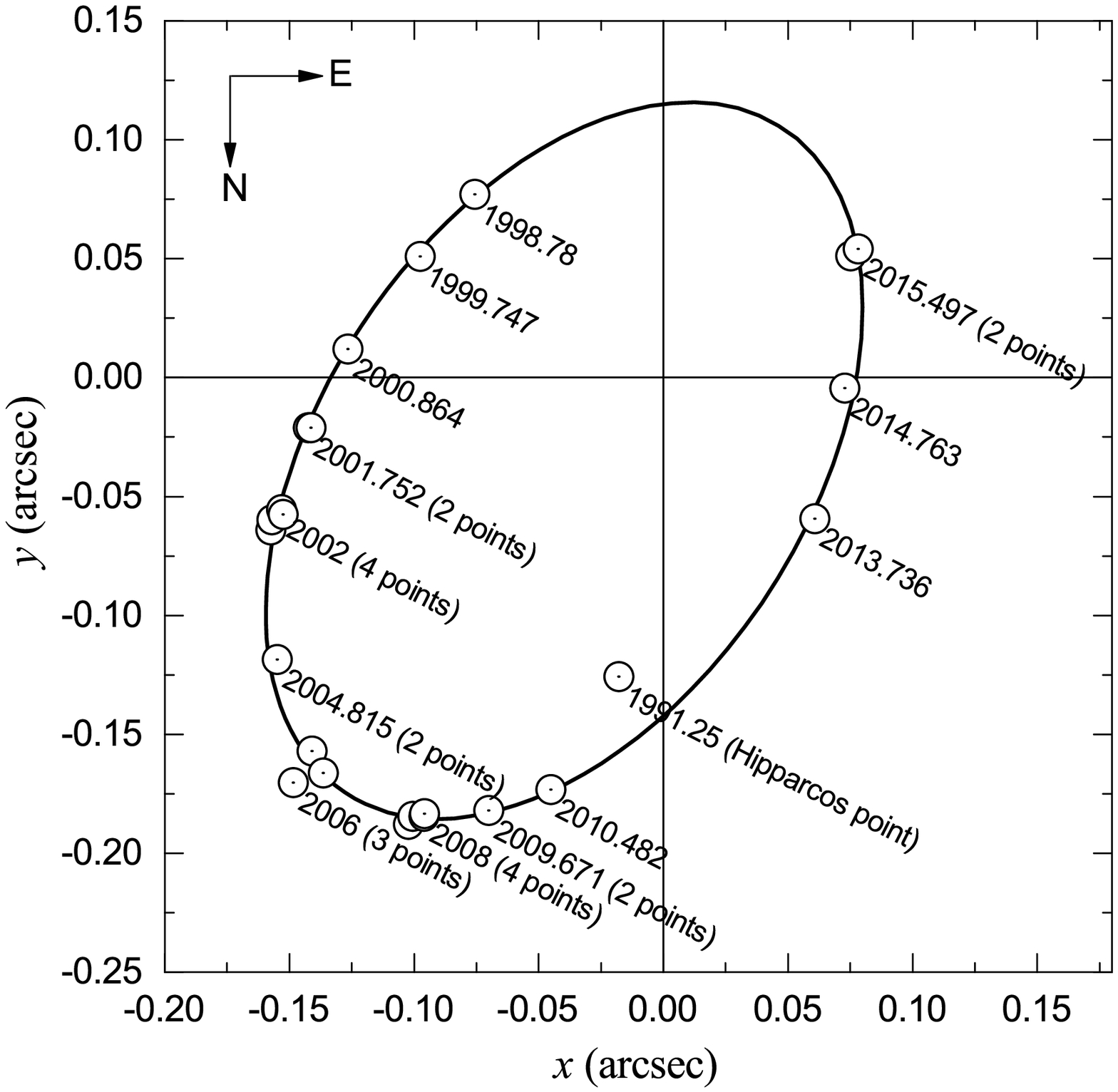}
	\caption{The relative orbit of the binary system HIP 105947  constructed using the relative position measurements from the Fourth Catalog of Interferometric Measurements of Binary Stars. The origin point represents the position of the primary component.
	} \label{figor}
\end{figure}

The results of the dynamical analysis and orbital solutions of HIP\,105947 are listed in Table~\ref{taba}, and the orbit is shown in Figure~\ref{figor}.

\subsection{Masses} \label{2.1}

As formerly mentioned that the mass is an intrinsically crucial parameter because the mass of the star determines its present and future, i.e. the birth, life and death of a star. As a result, using the estimated orbital solutions for the system in Table~\ref{taba}, period and semi-major axis, combined with
the revised \textit{Hipparcos} parallax~\citep{2007A&A...474..653V} for the system, led us to calculate
the total mass along with its error for the system according to Kepler's third law:
\begin{eqnarray}
\label{eq31}
\ M_{Tot.}=\ M_A +M_B=(\frac{a^3}{\pi^3P^2})\ M_\odot
\end{eqnarray}
\begin{eqnarray}
\label{eq32}
\frac{\sigma_M }{M} =\sqrt{9(\frac{\sigma_\pi}{\pi})^2+9(\frac{\sigma_a}{a})^2+4(\frac{\sigma_P}{P})^2}
\end{eqnarray}
where P is the orbital period (in years), $\ M_{A}$ and $\ M_{B}$ are the masses (in
solar mass), $a^{''}$ and $\pi$ are the semi-major axis and the \textit{Hipparcos} parallax
(both in arcsec), respectively. Using Equation~\ref{eq31}, we obtained the  mass-sum for HIP\,105947 as $2.06\pm0.36\,\rm M_\odot$.

Comparing this result with the estimated masses using Al-Wardat's method by plotting  the positions of the components on the H-R diagram (see~Table~\ref{tablef1} and Figure~\ref{evolall}), we find a good coincidence between them. This means that the analysis in both methods as well as combination of two methods together was successful, and consequently obtained the precise physical and geometrical parameters of the studied CVBS.
\begin{table*}[h]
	\small
	\begin{center}
		\caption{The estimated parameters of the individual components of  HIP\,105947.}
		\label{tablef1}
		\begin{tabular}{cccc}
			\noalign{\smallskip}
			\hline\hline
			&       	&\multicolumn{2}{c}{HIP 105947}  \\
			\cline{3-4}
			\noalign{\smallskip}
			Parameters & Units	& HIP 105947 A & HIP 105947 B  \\
			\hline
			\noalign{\smallskip}
			$\rm T_{\rm eff}$ {$\pm$ $\sigma_{\rm T_{\rm eff}}$}& [~K~] & $6230\pm80$ & $5470\pm80$ \\
			R {$\pm$ $\sigma_{\rm R}$}  & [R$_{\odot}$] & $1.443\pm0.06$ & $0.989\pm0.05$\\
			$\log\rm g$ {$\pm$ $\sigma_{\rm log g}$} & [cgs] & $4.30\pm0.07$ & $4.45\pm0.08$ \\
			$\rm L $ {$\pm$ $\sigma_{\rm L}$} & [$\rm L_\odot$] & $2.816\pm0.30 $  & $0.786\pm0.10$ \\
			$\rm M_{bol}$ {$\pm$ $\sigma_{\rm M_{bol}}$} & [mag] &  $3.63\pm0.08$ & $5.01\pm0.09$ \\
			$\rm M_{V}$ {$\pm$ $\sigma_{\rm M_{V}}$} & [mag] & $3.98\pm0.13$ & $5.51\pm0.14$ \\
			M $^{a}$ {$\pm$ $\sigma_{\rm M}$} & [$\rm M_{\odot}$]&  $1.21 \pm0.21$ & $0.89 \pm0.15$\\
			Sp. Type$^{b}$ & &  F8 & G7.5  \\
			\hline
			\multicolumn{1}{c}{Parallax $^{c}$  }
			& [mas]&  \multicolumn{2}{c}{$17.48 \pm 1.02 $}\\				
			\multicolumn{1}{c}{($\rm M_A+M_B $)$^{d}$}
			& [$\rm M_{\odot}$] &  \multicolumn{2}{c}{$2.06\pm0.36$}\\
			\multicolumn{1}{c}{Age $^{e}$}
			& [Gyr]& \multicolumn{2}{c}{ $2.75\pm 0.06$}\\
			\hline\hline
			\noalign{\smallskip}
		\end{tabular}\\
		$^{a}${Depending on the evolutionary tracks of masses ( 0.8, 0.9, ...., 1.3 $M_\odot$)\\ of~\citep{2000yCat..41410371G} (Figure~\ref{evolall}}).\\
		$^{b}${By using the tables of \cite{1992adps.book.....L,2005oasp.book.....G}} and depending on $\rm M_{V}-\rm S_{\rm P}$ relation.\\
		$^{c}${The revised \textit{Hipprcoas} parallax of $ 57.21\pm0.003$ pc for HIP 105947 ~\cite{2007A&A...474..653V}}.\\
		$^{d}${Depending on the orbital solutions and Tokovinin's method \cite{1992ASPC...32..573T}}.\\
		$^{e}${Depending on the the isochrones for low- and intermediate-mass stars of
			different \\metallicities of~\citep{2000A&AS..141..371G} (Figures~\ref{evolall1}\&~\ref{evolall12}}).	
	\end{center}
\end{table*}	
\begin{table*}[h]
	\begin{center}
		\caption{Orbital parameters solutions and total masses formerly published for the HIP\,105947 system, for comparison with this work .}
		\label{taba}
		\begin{tabular}{ccccc}
			\noalign{\smallskip}
			\hline\hline
			\noalign{\smallskip}
			&	&\multicolumn{3}{c}{System HIP 105947} \\
			\cline{3-5}
			\noalign{\smallskip}
			Parameters	& Units & \citep{2006AA...448..703B}& \citep{2010AJ....140..735M} & This work\\
			\hline
			$\rm P$ {$\pm$ $\sigma_{\rm P}$} & [yr]       & $18.79\pm 0.57$ & $20.3\pm 4.9$
			&$20.78\pm 0.06$\\
			$\rm T_0$ {$\pm$ $\sigma_{\rm T_0}$} & [yr]   & $1993.96 \pm 0.40$ & $1994.7 \pm 2.6$
			&$1994.94 \pm 0.06$\\
			$\rm e$ { $\pm$  $\sigma_{\rm e}$}  & -  & $0.354 \pm 0.010$ & $0.37 \pm 0.25$
			&$0.347 \pm 0.001$\\
			$\rm a $ { $\pm$ $\sigma_{\rm a}$}& [arcsec] & $0.168\pm 0.003 $ & $0.171\pm 0.018 $
			&$0.168\pm 0.0004 $\\
			$\rm i $ { $\pm$ $\sigma_{\rm i}$} & [deg]    & $ 57.1\pm 2.1$  &  $ 54.5\pm 3.9$
			&$ 51.78\pm 0.27$\\
			$\rm \omega$ {$\pm$ $\sigma_{\rm \omega}$}  & [deg] & $149.1\pm 1.5$  &  $140.0\pm 14.0$
			&$148.00\pm 0.72$\\
			$\rm \Omega$ {$\pm$ $\sigma_{\rm \Omega}$} & [deg] & $129.7 \pm 6.4$   & $154.0 \pm 12.0$
			&$152.41\pm 0.30$\\
			$\rm M_T ${$\pm$ $\sigma_{\rm M}$} & [M$_\odot$] & $2.68\pm0.58$  & $2.30\pm2.2$
			&$2.06\pm0.36$\\
			$\pi_{Hip}$ {$\pm$ $\sigma_{\pi_{Hip}}$} & [mas]   & $ 17.11\pm1.13$ $^{a}$ & $ 17.48\pm1.02$ $^{b}$
			&$ 17.48\pm1.02$ $^{b}$\\
			\hline\hline
			\noalign{\smallskip}
		\end{tabular}\\
		$^{a}${The old \textit{Hipprcoas} parallax} \cite{1997yCat.1239....0E},
		$^{b}$ {The revised \textit{Hipprcoas} parallax ~\cite{2007A&A...474..653V}}
	\end{center}
\end{table*}

\section {Results and discussion}

The results of  combining  Al-Wardat's complex method for analysing CVBSs with Tokovinin's dynamical method in analysing  the CVBS,  HIP\,105947, are explicitly presented.

Table~\ref{tablef1} shows the final physical and geometrical parameters of the individual components for the system. Figure~\ref{aa00} shows the best-fitting  between the entire synthetic and observational SED along with the  SED for each individual component presented for the first time.

Table~\ref{synth3} shows a very good coincidence  between the  synthetic magnitudes and colour indices with the observational ones for the system. Without a doubt, this gives lucid and powerful indication for the reliability of the  estimated physical and geometrical parameters of the individual components for the system as shown in Table~\ref{tablef1}.

Table~\ref{taba} summarizes the ultimate results of our orbital solutions with former studies for the system, and the orbit is plotted  in Figure~\ref{figor}. In addition to that, Table~\ref{po1} lists the residuals $\triangle\theta $ and $\triangle\rho $ and rms for HIP\,105947. The results of the orbital solutions show a slight modifications in the orbital parameters. The rms for our orbital solutions are $0.60\degr $ and $0\arcsec.0015$ for HIP\,105947.

\begin{figure}
	\centering
	\includegraphics[angle=0,width=12cm]{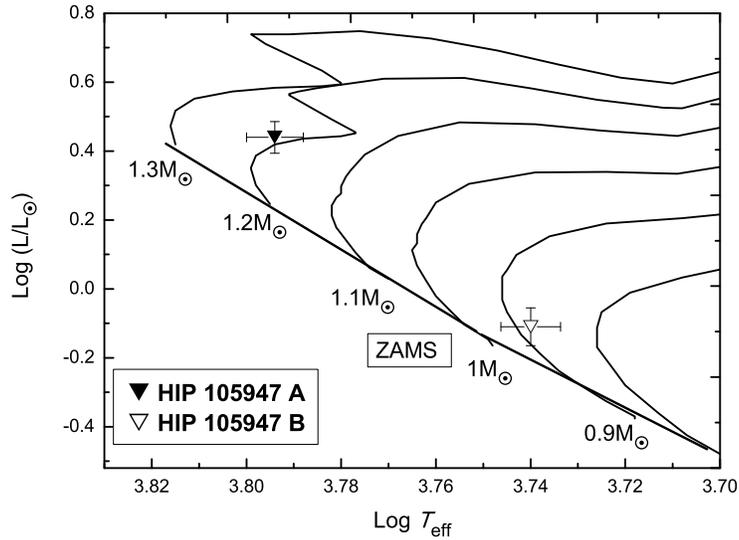}
	\caption{The evolutionary tracks of both components of HIP\,105947 on the H-R diagram of masses ( 0.8, 0.9, ...., 1.3 $\rm\,M_\odot$). The evolutionary tracks were taken from~\cite{2000yCat..41410371G}. }
	\label{evolall}
\end{figure}

The stellar masses for the system were calculated using two independent methods; Al-Warda's complex method for analysing CVBSs and Tokovinin's dynamical method for orbital solutions of binary stars. The former gives $2.10\pm0.19\,\rm M_\odot$ for HIP\,105947 (see Figure~\ref{evolall}\& Table~\ref{tablef1}), while the latter gives $2.06\pm0.36\,\rm M_\odot$ for HIP\,105947 (see Table~\ref{taba}). Both methods used the revised \textit{Hipparcos} parallax ~\cite{2007A&A...474..653V}.

A fast look and comparison with old orbital parameters for the system shows that for HIP\,105947, ~\cite{2006AA...448..703B} used the old \textit{Hipparcos} parallax~\cite{1997yCat.1239....0E} to estimate the masses, whereas~\cite{2010AJ....140..735M} used the revised \textit{Hipparcos} parallax but their eccentricity $\rm e$ and sem-major axis $\rm a$ were to a certain extent high unlike~\cite{2006AA...448..703B} and ours. Of course, these discrepancies are due to the availability of new observational measurements given in the \textit{Fourth Catalog of Interferometric Measurements of Binary Stars}.

Table~\ref{tab41} lists the  results of the  synthetic photometry of the entire system and individual components of  HIP 105947, in different photometrical systems (Johnson: $U$, $B$, $ V$, $R$, $U-B$, $B-V$, $V-R$; Str\"{o}mgren: $u$, $v$, $b$,
$y$, $u-v$, $v-b$, $b-y$ and Tycho: $B_{T}$, $ V_{T}$, $B_{T}-V_{T}$).

\begin{figure}
	\centering
	\includegraphics[angle=0,width=12cm]{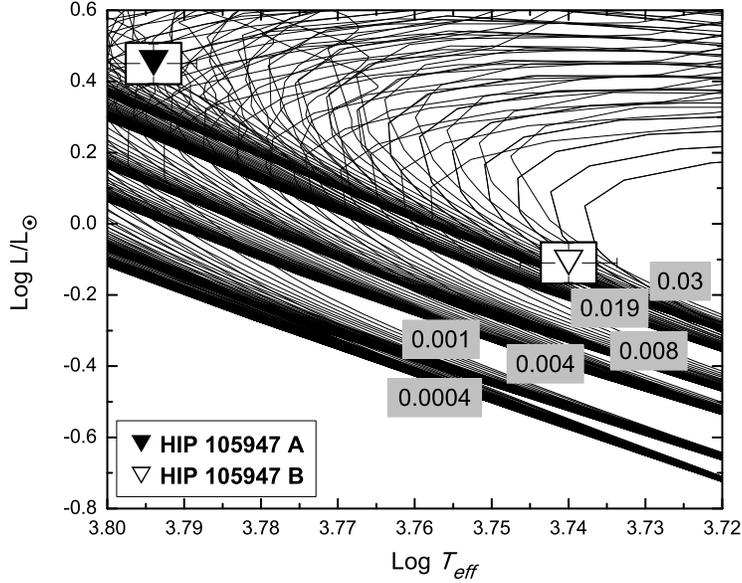}
	\caption{ The isochrones for both components of HIP\,105947  on the H-R diagram for low- and intermediate-mass stars of
		different metallicities. The isochrones were taken from~\cite{2000A&AS..141..371G}. }
	\label{evolall1}
\end{figure}

\begin{figure}
	\centering
	\includegraphics[angle=0,width=12cm]{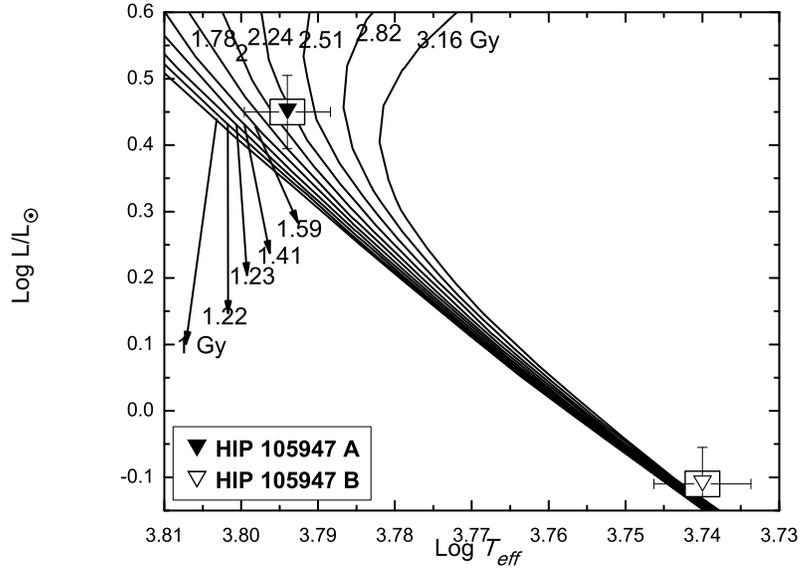}
	\caption{ The isochrones for both components of HIP\,105947 on the H-R diagram for low- and intermediate-mass: from $0.15 $ to $7 \rm\,M_\odot$ , and for the compositions [Z=0.019, Y=0.273] stars of different metallicities. The isochrones were taken from~\cite{2000A&AS..141..371G}. }
	\label{evolall12}
\end{figure}

The positions of the individual components of the system are shown on the evolutionary tracks of~\citep{2000yCat..41410371G} (Figure~\ref{evolall}). The Figure
shows that all components belong to the main sequence stars but the primary and secondary components of the system HIP 105947 are totally similar to the primary and
secondary components of the double-lined spectroscopic binary system HD 22128~\cite{2013MNRAS.433.3336F},
which are well-known as zero-age main-sequence. Therefore,
this led us to a broad suggestion that both components of the system HIP 105947
are  zero-age main-sequence stars. The analysis of the system  HIP\,105947 shows that both of its components are of age around $2.75\pm0.06$\,Gyr and solar composition [Z = 0.019, Y = 0.273] \citep{2000A&AS..141..371G} (see Figures~\ref{evolall1}\&~\ref{evolall12}).

\begin{table}
	\small
	\begin{center}
		\caption{Comparison between the observational and synthetic
			magnitudes and colours indices for both systems.} \label{synth3}
		\begin{tabular}{ccc}
			\noalign{\smallskip}
			\hline\hline
			\noalign{\smallskip}
			&\multicolumn{2}{c}{HIP 105947}\\
			\cline{2-3}
			\noalign{\smallskip}
			Filter	& Observed $^a$ & Synthetic$^b$(This work) \\
			& ($\rm mag$) & ($\rm mag$) \\
			
			\hline
			\noalign{\smallskip}
			$V_{J}$ & $7.53$ & $7.53\pm0.03$\\
			$B_J$& $8.10\pm0.02$ & $8.13\pm0.03$  \\
			$B_T$  & $8.24\pm0.02$   &$8.27\pm0.03$\\
			$V_T$  & $7.59\pm0.01$   &$7.60\pm0.03$\\
			$(B-V)_{J}$&$ 0.60\pm0.02$ &$ 0.60\pm0.03$\\
			$(u-v)_{S}$&$ 0.95\pm0.015$ &$ 0.95\pm0.03$\\
			$(v-b)_{S}$&$ 0.56\pm0.003$ &$ 0.59\pm0.03$\\
			$(b-y)_{S}$&$ 0.37\pm0.000$ &$ 0.37\pm0.03$\\
			$\triangle\rm m$  &$ 1.53^{c}\pm0.07$  &$ 1.53^{d}\pm0.05$\\
			\hline\hline \noalign{\smallskip}
		\end{tabular}\\
		Notes.\\
		$^a$ A real observations (see Table~\ref{tab1})\\
		$^b$ Synthetic work of the HIP 105947 system by using Interactive Data Language (IDL) program (see Table~\ref{tab41})\\
		$^c$ As the average of all $\triangle\rm m$ measurements under the speckle filters V-band only 534 -562 nm (see Table~\ref{tab22})\\
		$^d$ The magnitude difference between the A and B component ($\triangle\rm m$ = $V^{B}_{J}$-$V^{A}_{J}$) (see Table~\ref{tab41}) \\
	\end{center}
\end{table}
\section{Conclusions}
Using Al-Wardat's method for analysing CVBSs which employs  Kurucz {\fontfamily{cmtt}\selectfont ATLAS9} line-blanketed plane-parallel model atmospheres in constructing the entire and individual synthetic SEDs for a binary system, along with Tokovinin's dynamical method for orbital solutions of binary stars, we were able to estimate the complete set of physical and geometrical parameters of the Main-Sequence CVBS, HIP\,105947.

The calculated entire and individual synthetic magnitudes and colour indices of the system for different photometrical systems such as Johnson-Cousin UBV R, Str\"{o}mgren uvby and Tycho BV are introduced. Revised orbits
and orbital parameters for the system are introduced and compared with previous studies.

The ideal positions of the  components of the system are shown with a broad way on the evolutionary tracks and isochrones. The spectral types of the components of HIP\,105947 are catalogued
 as F8 and G7.5 for the primary and secondary components of the system, respectively with an age of $2.75\pm 0.06$\,Gyr.

Depending on the estimated physical and geometrical parameters of the system, the fragmentation process for the formation of such system is the most likely one, where the rotating disc around an incipient central protostar in case of continuing infall and the hierarchical fragmentation during rotational collapse are the main  mechanisms  in producing binaries and multiple systems, as pointed out by ~\cite{1994MNRAS.269..837B} and ~\cite{2001IAUS..200.....Z}.

\begin{acknowledgements}
	This research has made use of SAO/NASA, SIMBAD database, Fourth Catalog of Interferometric Measurements of Binary Stars, IPAC data systems, ORBIT code and CHORIZOS code of photometric and spectrophotometric data analysis. Suhail Masda would like to thank Human Development Fund for the scholarship and Hadhramout University in Yemen for perpetual support.
\end{acknowledgements}

\newpage

\end{document}